# Enabling Strong Database Integrity using Trusted Execution Environments

**Kai Mast**
Cornell University

**Lequn Chen**
Shanghai Jiao Tong University

**Emin Gün Sirer**
Cornell University

March 2, 2018

## Abstract

Many applications require the immutable and consistent sharing of data across organizational boundaries. Because conventional datastores cannot provide this functionality, blockchains have been proposed as one possible solution. Yet public blockchains are energy inefficient, hard to scale and suffer from limited throughput and high latencies, while permissioned blockchains depend on specially designated nodes, potentially leak meta-information, and also suffer from scale and performance bottlenecks.

This paper presents CreDB, a datastore that provides blockchain-like guarantees of integrity using trusted execution environments. CreDB employs four novel mechanisms to support a new class of applications. First, it creates a permanent record of every transaction, known as a witness, that clients can then use not only to audit the database but to prove to third parties that desired actions took place. Second, it associates with every object an inseparable and inviolable policy, which not only performs access control but enables the datastore to implement state machines whose behavior is amenable to analysis. Third, timeline inspection allows authorized parties to inspect and reason about the history of changes made to the data. Finally, CreDB provides a protected function evaluation mechanism that allows integrity-protected computation over private data. The paper describes these mechanisms, and the applications they collectively enable, in detail. We have fully implemented a prototype of CreDB on Intel SGX. Evaluation shows that CreDB can serve as a drop-in replacement for other NoSQL stores, such as MongoDB while providing stronger integrity guarantees.

## 1 Introduction

Many high-value applications require the reliable and immutable storage of data across multiple distrusting parties [51, 17, 56]. These applications are characterized by integrity requirements wherein each party must abide by pre-defined policies. Conventional databases cannot live up to this challenge, as they require full trust in the database application and host operating system.

Blockchains have recently emerged as a potential platform to address the needs of these applications. Public blockchains [38, 52], based on Nakamoto consensus, maintain an immutable log of events distributed across all participants of the system. As a result, they are energy inefficient, hard to scale and suffer from limited throughput and high latencies [15]. Further, due to their open and distributed setting, they cannot be used to store private or confidential data. Permissioned blockchains [10, 3, 2] employ a committee consensus protocol [11, 28, 23] to maintain the log and append updates in an orderly fashion. Changes can only be made if a specified quorum of the committee agrees to do so. This approach necessarily requires specially designated committee nodes, exposes at least meta-information to those nodes, and is limited in performance by bottlenecks in the quorum. These efforts were preceded by earlier work on accountability systems [27, 53], which ensure integrity by allowing clients to audit the log with respect to states they have observed previously. But accountability mechanisms can only enforce fork consistency [35, 18], a weaker security property than strong consistency.

This work presents CreDB, a novel datastore that provides the integrity guarantees of blockchains, using much more efficient and scalable techniques backed by secure hardware. In a nutshell, CreDB provides an append-only log [13] of partially-ordered states. Past states cannot be changed and new states can only be appended to the log. Building on this foundation, CreDB nodes can operate independently, as a "blockchain of one," or as part of a network of nodes that can share designated data items and invoke computations on each other. Each node in the system runs in a *trusted execution environment (TEE)*, provided by the system's hardware. The usage of TEEs enables nodes to trust another participant's computation without trusting the administrator of that system. Because each service can run their own database node backed by a TEE, the throughput scales with the number of nodes in the system.

CreDB nodes issue witnesses, which are permanent and tamper-proof certificates of the state of the system. Further, they are independently verifiable, i.e. verification does not dependent on a specific node in the system. Witnesses can be used to establish facts about the datastore, such as the instantaneous contents of objects, the existence of certain data or past transactions, and ordering of transactions. This enables even untrusted applications, backed by CreDB, to provide proofs of their correct operation to third parties. Because witnesses are free-standing, they enable parties who are not direct clients of the database to verify crucial aspects of the database's operation.

CreDB enables every object to be coupled with an associated semantic security policy. Because policies are enforced by TEE-backed software, they are inseparable from their associated data. And because they are written in a Turing-complete language, can express rich, object-specific policies. And because the semantic security policies are encoded symbolically as abstract syntax trees, they are amenable to analysis by third parties. Coupled with witnesses, these techniques enable a third party to inspect the policy associated with an object and thus establish trust in the future behavior of that object. These capabilities can be used to build *smart contracts* on CreDB.

Finally, CreDB provides a *protected function evaluation* mechanism that enables clients to compute functions over remote private data, which in turn generate witnesses carrying the result. For the party issuing the function call, the witness yields a verifiable, portable certificate that the function has been executed, with integrity, on the specified data, with the attached result. The primary use of this functionality is to compute a vetted function over private data without revealing the input data to the remote party. For security purposes, the holder of the data retains full control over what can be done with the data, and both parties, the invoker, and the data holder must agree on which functions can be executed.

While there has been much past work on high-integrity data storage and data processing systems, to our knowledge, no datastores exist that combine these three synergistic features. Ryoan [21] and Opaque [55] have examined functions can be executed across private data. Guardat [49] reduces the attack surface of a system by enforcing data policies on the storage layer. Further, Cipherbase [4] and EnclaveDB [41] enforce security properties using trusted hardware. Finally, past systems have explored the use of trusted hardware to enhance audit mechanisms [13, 25]. CreDB takes these concepts and provides one holistic approach to secure and tamper-proof data storage, built using modern trusted hardware.

The rest of this paper is structured as follows. The next section provides the overarching data and computation model for CreDB (Section 2). Section 3 details how this model was implemented using Intel SGX. The following two sections evaluate a full prototype of the system. The evaluation contains a qualitative part, explaining how to implement several sample applications (Section 4), and a quantitative part, describing the impact our execution environment has on the performance of the system (Section 5). We show that, depending on the workload, CreDB can process up to 50k operations per second on a single machine. Further, we show that it can process about 500 transactions per second on the TPC-C benchmark.

## 2 The CreDB Data Store

At a high level, every CreDB node implements a secure database using a trusted execution environment (TEE) that clients, as well as other nodes, can connect to. Each database instance has its own timeline, datastore, and set of connected nodes. A CreDB node connects to other nodes in order to create a network across which data can be shared, and functions can be invoked, securely. Clients connect through one or multiple CreDB nodes and do not need special hardware support. This enables porting legacy database applications to this new abstraction and provide them with stronger integrity guarantees. Nodes and clients rely on a public attestation service to ensure integrity and authenticity of database nodes. Attestation services provide a public-key infrastructure to ensure the authenticity of parties. No private data is transferred to the attestation service.

Applications are written against the CreDB API, which is a superset of a traditional key-value API, and connect to a CreDB

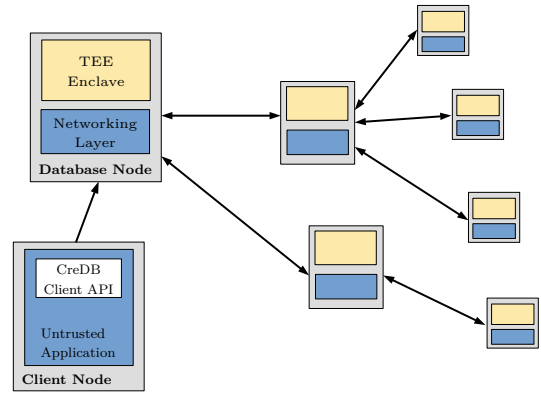

Figure 1: Sketched layout of the system. A network of database nodes, each running in their own TEE. Further, client nodes connect user-facing applications to the network. All nodes are interconnected through encrypted channels.

node of their choice. The most notable additions are *timeline inspection*, *secure semantic policies*, *witness generation*, and *protected function evaluation*, which we discuss below after we provide the basic object and event model on which the system is based.

### 2.1 Assumptions and Attack Model

Following previous work [6] we assume a very strong attack model: an adversary might have root access to the database server. This includes full control over the scheduler, the file system, and network communication. The attacker may tamper with the hardware, except for the CPU itself.

We further assume that clients may mistrust each other. This means clients need assurance that data can only be modified by parties they specify. Further, they demand control over what information is leaked to other parties, including the database administrator.

Replication and crash recovery are out of the scope of this work, however, the presented model can be extended to be fault-tolerant. This paper discusses such mechanisms in Section 6.

### 2.2 Objects and Events

CreDB exposes a flexible object model that accommodates unstructured, as well as structured, data. Objects are collections of attribute-value pairs, where attributes can have types such as lists, dictionaries, binary data, and primitive values, which consist of integers, floating point numbers, and strings. Binary data can contain executable code representing stored procedures. Each object is associated with a specific *collection* (similar to tables in relational datastores).

The datastore maintains a partially-ordered log of *events*, each relating to one or more *objects*. Events record the creation, update, or deletion of an object. Events store the new value of all updated objects. In the case of a deletion, the new value is a tombstone entry ⊥. Figure 2 gives an example of the lifetime of an object x, where each modification creates a state transition and a new event that holds the most recent value. Events relating to the same object are arranged in a total order to guarantee linearizability [20]. Further, in case events are created by a transaction that spans multiple objects, an event may also capture the dependencies between versions of different objects. For instance, Figure 3 shows an event from a transaction that reads from object "foo" and then writes to object "bar". Crucially, events that are unrelated are not ordered with respect to each other.

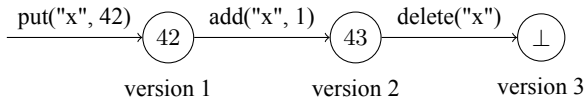

Figure 2: The lifetime of an object mapped to a sequence of events. Because CreDB guarantees strong consistency this order must be total. ⊥ represents a deleted object (i.e. *"tombstone"* entry)

## 2.3 Timeline Inspection

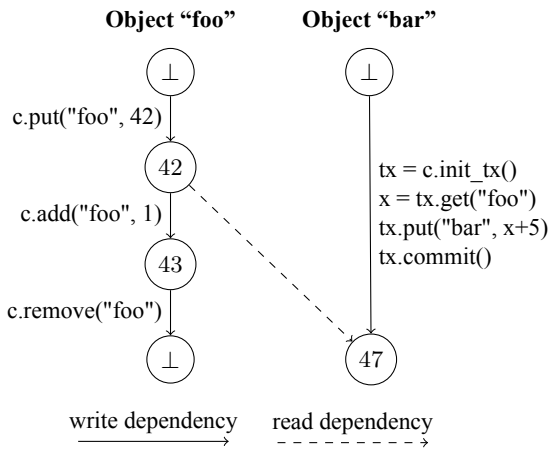

Figure 3: A transaction encoding its causal dependency, which stems from a read it performed, into the timeline. The log encodes how two concurrent clients affect the objects.

Unlike a traditional key-value store, CreDB enables historical retrievals from any previous point in time. The desire for such rich semantics has been underscored by the interest in both blockchains and bitemporal databases [44]. CreDB thus implements an eidetic database that stores the set of all modifications to all objects, which is essential for analysis and audit and produces an API for inspecting historical values.

The CreDB API enables clients to retrieve previous versions of an object using the get_version and get_history calls. get_version takes an object key and version id and returns that object's state at that specific point in the timeline as well as the corresponding event id. Similarly, get_history takes an object key, and returns a sequence of all versions of an object and their event identifiers. An event id can then be used to retrieve more information about the context that an object was modified in. In particular, if the object was created as part of a transaction, it allows to retrieve the read and write sets of said transaction. This is enabled through the get_read_set and get_write_set. For all events, we can further query the party that issued the modification using the get_op_source call.

Each event identifier corresponds to a tuple of an object key and a version number. While version numbers of the same object are totally ordered, event identifiers are not required to be. This enables the datastore to support multiple concurrent updates and obviates the need to have a single, totally-ordered log, which would hamper performance. This design decision is coupled with our decision to have a partial order between unrelated events. Thus, our API is only guaranteed to answer questions of the form "what was X's value when I updated Y?" if X was in the read set of the transaction that updated Y.

Figure 3 illustrates how a read-dependency is encoded. When inspecting the event that sets "bar" to 47, the API will return a reference to the value of "foo" when it was read by the transaction. The immutable log ensures that this timeline is final and can be reasoned about safely. The database can then answer questions such as "who has updated X since it was first created?". For instance, in the example from Figure 3 calling get_history on object "foo" would return a sequence of [42, 43] and their corresponding event identifiers.

## 2.4 Secure Semantic Policies

Secure semantic policies enable applications to associate application-specific constraints with an object. These policies are inseparable from the object to which they belong and inviolable even by the principal controlling the database instance. To access the database, a user must necessarily go through the CreDB secure semantic policy enforcement engine mandated by the TEE. Thus, even an attacker who takes over the database cannot subvert the access policies associated with objects. In case of accessing a previous version of the object, that version's policy and state will be used to make an access control decision.

Each CreDB node maintains a registry of identities, which can be leveraged by policies to make an access decision. Identities are tuples consisting of a human-readable name and a public key. This registry is used to prevent man-in-the-middle and impersonation attacks. We assume a public key infrastructure (PKI). When a previously unknown party connects to a node, it queries the PKI to gather information about said node.

Identities are inseparable from the associated authenticated communication channel. In particular, nodes cannot change their identity after a connection has been set up. The rather complex authentication and attestation mechanisms then only has to be done once, when setting up the channel. Policies and stored procedures can always rely on the authenticity of the referenced identities.

Policies are specified at the time of an object's creation and can be modified after the fact only if the policy permits it. And changes to a policy are stored in the object's timeline just like changes to all other fields of the object. Accesses to an object's value in the timeline lead to the evaluation of the object's policy at that point in time. In order to enable this, CreDB requires object policies to be idempotent, i.e. they may not have any side effects or reference the state of other objects. CreDB enforces this by restricting the policy interpreter from performing such operations.

Policies use the current state of the object, as well as information about the attempted operation, to make access control decisions. Similar to conventional stored procedures policies have access to several modules that hold information about the attempted operation and the object to be accessed. Two modules, in particular, are only available to secure semantic policies: First, op_info contains information about the operation itself, such as the kind of operation and the proposed change. Second, op_context enables to retrieve information about the issuing party, such as their identity. A full enumeration of available modules is given in Table 1.

| Name | Description | PFE | SSP |
|---|---|---|---|
| db | Interact with the database as a client would do | Yes | No |
| self | If the function is part of an object, this module provides access to that object | Yes | Yes |
| rand | Bindings to the random number generators provided by the TEE | Yes | No |
| op_info | Information about the attempted operation | No | Yes |
| op_context | Information about the invokation of the operation | No | Yes |

Table 1: Python Modules available to Server-Side Programs. Some are only accessible by Protected Function Evaluation (PFE), some only by Secure Semantic Policies (SSP), others by both.

Policies are then represented as a single function that returns a boolean. Figure 1 illustrates a simple policy that restricts access to an object to a pre-defined set of users. Here, aside from the policy itself, an object contains two more fields: authorized_users specifies who is allowed to access the data and data holds the confidential data itself. More complex policies are enabled due to the fact that SSPs can make use of timeline inspection. In particular, they can examine the object's timeline, for instance, to determine how often an object has been modified, in total or by a specific client, or check the object's history against a predicate.

To enable richer application semantics, CreDB further allows associating a policy with a collection. Such collection policies may, for example, specify who can create or modify any object in the collection. Further, they can enforce a schema on the data, by rejecting all updates that miss required fields or contain fields in an invalid format. Collection policies thus allow to break down application logic into multiple concurrent objects without sacri-

Listing 1: A policy limiting authorized users to read the data field.

```python
import self
from op_info import target
from op_context import source

if target != "data";
    # only the data field can be accessed
    return False

users = self.get('authorized_users')
return source in users
```

ficing integrity.

## 2.5 Witnesses

*Witnesses* are a permanent, external, and free-standing record of the database state. They encode a set of events with respect to their position in the timeline and are signed by the private key of the CreDB instance. Because of their free-standing nature, witnesses are useful even beyond the existence of the issuing service.

A witness is comprised of a list of events, as well as their location in the timeline and potential causal dependencies between each other. Witnesses are generated automatically in response to every transaction on the database and passed along with the transaction's result to the calling party. The set of events inside a witness then corresponds to the state of the object at the time the transaction read or updated them.

Witnesses enable auditing applications that use CreDB as their datastore. Applications generate witnesses by issuing transactions to the datastore, which can then be shown to other parties as a proof of action. For example, witnesses can serve as a payment receipt or a proof of ownership for a certain asset. Further, witnesses can be used to prove that an update was not made using false assumptions by capturing the read set of a transaction at the time of commit. For example, a bank clerk who wants to show that they observed an irregularity in a bank account before declining a credit request can refer to a specific point in an object's timeline by using a witness.

The CreDB API supports three key actions on witnesses: (1) verifying them for authenticity, (2) examining witness contents, and (3) ordering witnesses with respect to each other. Witness verification ensures that a witness is authentic, that is it was created by a TEE running CreDB. This can be achieved by checking the witness against the public key of the principal provided by the attestation service. Witness verification does not explicitly guarantee freshness, but a witness can include a timestamp signed by a trusted time source to indicate the time at which it was signed. In applications where establishing an absolute time is not necessary, a relative order can be established using the timeline inspection primitives described below.

By allowing clients to extract the contents of witnesses, we enable local timeline inspection without the need to access the datastore itself. A local timeline can be generated from a single witness or a set of overlapping witnesses. The clients then have access to a timeline inspection API that provides similar function calls as those when interacting directly with the server.

Local timeline inspection then enables checking the order of events, the relative order of two witnesses, or specific events contained in the witnesses. A call to check_order can return: before, after or incomparable. The datastore may return incomparable when comparing two concurrent set of events operating on independent data items. In cases where it should be possible to always order two witnesses, a common object must be included, which will never yield incomparable. This approach enables CreDB to achieve high performance, as it does not impose a total order on all concurrent transactions and instead permits the timeline to be structured as a directed acyclic graph.

Together, these three actions enable application clients to audit application behavior without direct access to the datastore itself. Clients can reason about the correctness and trustworthiness of the application logic by collecting and verifying witnesses that are generated by the datastore and passed on by the application. For example, they can inspect which principals have access to their sensitive data by requesting a witness containing the data's access policy. A certified access policy can further provide a guarantee that the data is protected from a malicious application.

### 2.6 Protected Function Evaluation

Another key primitive supported by CreDB is *protected function evaluation (PFE)*. PFE enables parties to invoke a custom function on a remote node in a secure execution environment guarded by the TEE. This way, data protected by the TEE remains private to the trusted environment, and only the designated result of the function call is revealed to the caller.

Since computations on private data have the intended goal of retrieving some information extracted from that data, they need to be vetted to ensure that this leakage is permissible to all parties. CreDB employs two mechanisms to perform this vetting. First, prior to execution of a function, both the calling and the executing parties must approve the function. The executing party needs to ensure that no private data is leaked and that the function execution does not take up an unreasonable amount of resources. This can be done by checking the functions hash against a whitelist or by analyzing the AST of the function. Much past work concerns itself with the analysis of function properties, including for information leakage [16] and information flow [30], and is beyond the scope of our work. In addition, every single object retrieved during a PFE has its semantic security policy checked on every access.

After successful execution, the calling party receives a witness containing a function identifier and its result. CreDB identifies functions through the hash of their bytecode. The witness is signed by a persistent key associated with the CreDB instance of the executing party.

This design imposes minimal structure on CreDB witnesses. In particular, it deliberately leaves freshness guarantees up to applications – CreDB does not purport to provide a global clock or a total order of events. The critical observations behind this decision are threefold. First, no single notion of time can serve every application. Some applications may operate on a sub-microsecond granularity, which could entail inordinate overheads, while others keep track of events in a more coarse-grained manner. Second, even if there was a time granularity that one could pick for most applications, current technologies for providing a trusted time source into a secure execution environment provide much weaker guarantees than the TEE itself, because they rely on additional hardware outside of the CPU die [14]. Finally, it has been our experience that most applications can be implemented using simple happens-before relationships between the affected objects.

### 2.7 Summary and API

We outlined the four core features provided by CreDB, beyond conventional key-value storage. Together, these synergistic features make CreDB a high-integrity datastore which protects the data from unwanted access and provides strong accountability. Thus, CreDB can be described as a "blockchain of one," a self-standing blockchain that does not rely on expensive data replication or other consensus mechanisms.

Table 2 shows all operations supported by CreDB. In addition to executing standalone, operations can be executed in the context of a transaction. Until committed, transactions create a locally isolated view of the database and the attempted changes. These changes are only merged if no conflicts are detected on commit.

## 3 Implementation

We implemented a fully-functional prototype of the CreDB database. The prototype is built on top of the Intel SGX SDK, which provides a well-documented TEE with certain restrictions. In particular, the memory encryption mechanism that protects TEEs from unwanted access causes a severe limitation on memory that can be accessed efficiently. SGX hardware supports an encrypted page cache (EPC) that can be accessed through an in-hardware encrypted page cache module (EPCM) [14]. Currently, all SGX-enabled CPUs support an EPC of at most 128MB. This EPC also needs to hold program code and stack. From our experience, this results in an available heap size of roughly 90MB. While future generations of CPUs might increase the EPC size, it seems unlikely that such a size restriction disappear completely. Not only data that is held in heap memory, but also function arguments need to be encrypted and decrypted when crossing enclave boundaries, for example when a network message is passed from the untrusted part of the database to the enclave.

In order to overcome these performance limitations, our prototype design follows two goals. First, we aim to keep the overall memory consumption low, by using efficient and lightweight data structures. Second, the implementation minimizes the amount of memory that has to be moved in and out of the EPC. The latter is achieved through a custom paging mechanism described in the next section.

## 3.1 A concurrent append-only log

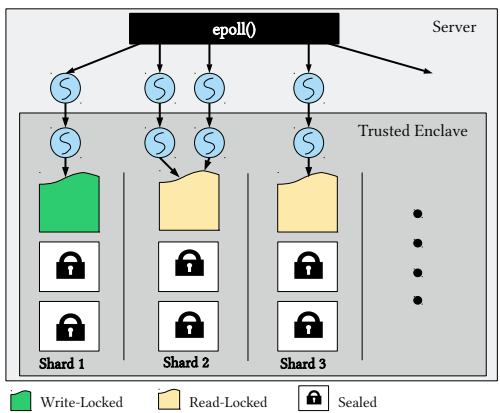

Figure 4: Implementation of a CreDB node: The keyspace is broken down into multiple shards. Each shard stores its events in a sequence of blocks.

CreDB nodes achieve high throughput by breaking down their keyspace into $n$ shards. Consistent hashing is used to assign a key range to each of the shards, where $n$ is set sufficiently large to exceed the number of hardware threads available. Each node then runs multiple concurrent threads, inside and outside the enclave, to efficiently leverage the sharded setup, visualized in Figure 4. The untrusted threads wait on a multi-threaded event loop built on top of epoll. Once an event is dispatched by an untrusted thread it is forwarded to a trusted thread. In our prototype, the ratio of trusted to untrusted threads is 1, such that once an untrusted thread has dispatched a new message, there is always a trusted thread available to process it. We set the number of threads to twice the number of available CPU cores. Threads then update the log by working with one or multiple shards. In order to achieve strong consistency, only one thread at a time can update the same shard. This is enforced through a read-write locking mechanism inside the TEE.

In order to enforce the append-only mechanism for the log, the timeline for each shard is sequenced into blocks. Only the newest block is then modified when a new event is inserted into the log, the other blocks are considered sealed. A primary index structure allows retrieving the most recent event associated with an object, without having to perform a linear scan across the log. Secondary indexes can be created to quickly react to more complex queries. CreDB's custom paging mechanism moves blocks in and out of the EPC. Previous work has shown that a reference count based mechanism can achieve a significant performance benefit compared to SGX native paging mechanism [39]. A dedicated, shard-aware, and CreDB-specific paging mechanism allows accommodating arbitrarily large data sizes in the enclave's fixed heap space. Blocks are only evicted if no threads are using them at the current point in time. Eviction is run whenever the number of blocks in EPC memory exceeds a certain size. Besides being stored on the local file system, encrypted blocks are also cached in the untrusted memory to retrieve them quickly.

A similar paging mechanism is also employed for primary and secondary indexes.

## 3.2 Ensuring Data Freshness

Because paging moves data out of the enclave, CreDB employs mechanisms to avoid stale data from being loaded back into the enclave. The enclave can easily detect corrupt data due the to the encryption scheme applied to the data blocks, however, a separate measure is required to detect stale data from being loaded. For each shard, only the most recent, i.e. *pending*, block is considered mutable. On each modification, the pending block will be saved to disk. Once a block has reached a specific size it will be marked sealed both in memory and on disk. The enclave then always keeps the pending block in memory and only allows to load sealed blocks.

Unlike data blocks, indexes are mutable and require a different mechanism to ensure their freshness. CreDB implements authenticated hash maps for this purpose. The hash map breaks down the keyspace into buckets where each bucket is essentially a linked list of key-value pairs. For each bucket, the hash map holds the identifier of the first page of that bucket as well as a version number, which is a simple integer. These references are always kept in memory. Each page then holds the identifier and a version number for their successor, if it exists. On each update, all version numbers and references of that bucket are updated. When loading a page the implementation can then check it's freshness by comparing version numbers.

## 3.3 Efficient Transaction Processing and Witness Generation

CreDB implements serializable transaction that yield witnesses. Transaction processing uses optimistic concurrency control [24] to prevent malicious parties from breaking liveness of the system. The party initiating the transaction sends a message containing all reads and a set of writes. In the *validation phase*, the node checks that policies grant all reads and modification made by the transaction. It further validates all reads made during the speculative execution on the client side. If no conflicts are detected, the node proceeds to the *write phase*, where it issues all updates to the log by creating new events. Further, events are annotated with causal dependencies using the transactions read set. Locks are only acquired during validation and write phase of the transaction, all of which can be executed on the server-side without leaving the trusted environment. Further, locks to shards are always acquired in the same order, to ensure no deadlocks are introduced during concurrent transactions.

We then extend the transaction primitive to enable accountability by generating witnesses after a successful commit. Clients can request the generation of a witness as a result of a transaction. In particular, we add a *certification phase* after the write phase. This step is optional, in order to allow clients that do not need self-standing witnesses to avoid additional overheads. The certification phase takes the read set, as well as the set of events created in the write phase. From these, it generates a digest and signs it using the enclaves private key. This way, only a single asymmetric cryptographic operation per transaction has to be executed by the CreDB node.

## 3.4 Peering Handshake

All communication between other CreDB nodes and clients require a secure, confidential, and authenticated channel in order to uphold all of the datastore's integrity guarantees. This is achieved by using a remote attestation. Remote attestation consists of a handshake that serves two purposes. First, it sets up an encrypted channel using a Diffie-Hellmann Key Exchange (DHKE) rooted in the TEE. Second, it verifies that the TEE is executing a non-modified version of the CreDB implementation. Previous work has described how such attestation channels can be set up, for example, to establish a secure payment channel [29].

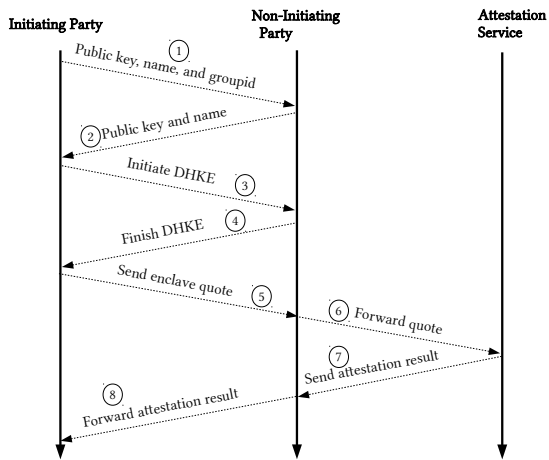

Figure 5: The attestation handshake establishes a secure channel between two parties. If both of them provide a trusted environment, it has to be executed in an identical fashion going the other way.

The attestation process can be broken down into eight steps, which are visualized in Figure 5. (1) The initiating party sends its Group ID (an SGX specific detail), public key (used to verify authenticity), and name (for easy addressing of the node). (2) The other party then responds by also sharing its public key and name. Parties can store known public keys and/or use public key infrastructures to ensure the authenticity of other parties, as described in Section 2.4. (3)(4) Once both parties have shared their public keys they can initiate a DHKE to generate a shared secret. (5) The initiating party then sends a *quote* of their enclave. A quote is generated from a snapshot of the enclaves content, including both data and loaded program code. It can thus be used to reliably identify the code running on the remote party. (6)(7) The other party forwards the quote to an attestation service. Attestation services are Intel-verified entities that can check the enclave's signature for validity. This step is needed to support revocation of keys of malicious CPUs. (8) Once the attestation service has verified the quote, the result is sent back to the initiating party. At this point, assuming the quote and signature were valid, a one-sided authenticated channel is setup up.

In order to set up a bidirectional channel between two CreDB nodes, this attestation handshake is performed once in each direction. There might be ways to shorten the bidirectional handshake process by avoiding redundant messages. However, the latency of remote attestation does not affect overall performance as the CreDB model assumes long-running connections between nodes.

## 3.5 Trusted Program Execution

Efficient program and policy evaluation are enabled by a lightweight Python interpreter that runs inside the TEE. Our implementation aims to find a middle ground between ease of use and efficiency. CreDB clients come with a compiler that can convert a subset of Python to a compressed abstract syntax tree (cAST). The cAST can then be stored as bytecode on a CreDB node. Programs run in isolation but are able to access other parts of the TEE using Python bindings that are shipped as part of the implementation.

The execution environment further protects nodes from malicious programs exhausting the EPC size. Programs can be executed with pre-defined *execution limits*: such as the amount of memory it can use up the most, the number of execution steps it is allowed to take, and the number of remote functions it is permitted to call. The node keeps track of the program memory consumption and amount of execution steps, if execution exceeds either limit, it is aborted. The result of the function call is then either an error code representing why the function was aborted or the result of the function itself.

Each execution environment is assigned its own userspace thread that can safely be suspended if needed. Programs run non-preemptively and either pause when waiting for a message from another node, or stop when they have exhausted execution limits. This achieves two purposes. First, hardware threads are not occupied by programs that are waiting for an I/O operation to finish. Second, execution of the system will not halt, even if more programs execute than the number of threads supported by the TEE. This overcomes another limitation of the current SGX generation, which is that TEEs can only execute up to a pre-defined limit of hardware threads.

## 4 Example Applications

CreDB provides a novel programming model for building high-integrity applications in a Byzantine environment. To demonstrate the practicality of this model, this section describes multiple applications. These applications leverage server-side program evaluation, both in the form of stored procedure and policies, to provide rich application logic on top of CreDB. Together with stored procedures, policies enable to implement state machines in form of an object. Stored procedures can define possible state transitions, while the policies restrict when and by whom these transitions can be invoked.

## 4.1 Multi-Tenant Datastores

Keeping data secure from unprivileged access is a common problem in enterprise datastores. Often a shared nothing architecture, where each tenant has their own distinct protected storage area, is not feasible as tenants might still want to exchange, or grant temporary access to, privileged data. For example, HIPAA compliant systems have strict rules on who can access the data but need the possibility to grant special rights to users in case of an emergency.

Listing 2: An access control policy

```
import self
from op_context import source_name
from op_info import is_modification

ACL_NONE  = 0
ACL_READ  = 1
ACL_WRITE = 2

if self.contains("acl." + source_name)
    val = self.get("acl." + source_name)

    if is_modification:
        return val & ACL_WRITE
    else:
        return val & ACL_READ

else:
    # Deny access if no entry in ACL
    return False
```

Access Control Lists (ACLs) are the state-of-the-art mechanism to enforce multi-user access rights on a set of objects. Each ACL contains a mapping from user to permissions. A permission can give users different levels of access, such as read-only, read/write, or execute. Permissions can then be dynamically changed by authorized users.

We demonstrate that an ACL mechanism can be implemented, and enforced, using CreDB's policy system. To achieve this we extend the policy from Listing 1 to associate each object in the datastore with its own ACL. Each object then holds an acl-field, which contains a mapping from users to permissions. The object's policy inspects the acl-field and decides whether to allow an access or not depending on its value. Further, users with the right privileges can modify the acl-field and, thus, change the permissions during runtime.

## 4.2 Mandatory Read Logging

CreDB does not log read accesses by default, but applications can enforce such a mechanism. We demonstrate a mechanism similar to one presented by Vahldiek et al. [49], that requires clients to create a log entry before accessing the data. This mechanism cannot track how often an object is read, or whether the read

Listing 3: A policy that enforces each read to be logged: Before data can be read prepare_call must be invoked.

```
{
  read_log: list := []

  data: dict := {}

  prepare_read: func :=
    from self import contains, version_no
    from op_contenxt import source_name

    log_entry = {source_name, version_no()}
    return append("read_log", log_entry)

  policy: func :=
    from self import contains, version_no
    from op_context import source_name
    from op_info import target, type

    if target == "read_log":
        # Only reads to the log
        return type == READ
    elif target == "prepare_read":
        return type == CALL
    elif target == "data":
        if type == READ:
          log_entry = {source_name, version_no()}
          return contains("read_log", log_entry)
        else:
          # It's a write
          # This will be logged by default
          return True
    else:
        # Everything else is disallowed
        return False
}
```

access was successful. However, it requires each read to be preceded by a log entry, which collects information about attempted reads to the data.

Listing 3 illustrates how CreDB can implement this policy. The object consists of the read log, the actual data field, as well as two functions policy and prepare_read. When a client wants to update the value of an object, it can just issue the write directly, as CreDB will take care of creating and logging a new version. However, before the client issues a read, it will need to invoke the prepare_read function, which creates a log entry with the name of the client and the current version number of the object. When the client, then, tries to read the data-field, the policy will check whether a corresponding log entry exists before granting access. The client's identity is authenticated by the mechanism described in Section 2.4. Note that the policy can be extended to include other access control mechanisms.

### 4.3 Checking Credit History

In order to make a decision on whether or not to grant a new loan, a credit issuer might check the customer's bank for their credit history. However, the bank might not want to reveal customers' credit history for both privacy and business reasons. Further, the customer might have multiple bank accounts that need to be checked. Currently, this problem is addressed through a third party credit score agency that is trusted by all parties. Such an approach might not always be feasible as it requires such a common trusted party to exist and usually creates additional fees and pose additional vulnerabilities.

CreDB's PFE mechanism natively accommodates such blind checks. We built a verified credit score checker and execute it on the bank's data, without getting access to the credit data itself. Further, it supports nested PFEs, i.e. PFEs invoking other PFEs during their execution. The function can be vetted to ensure that it returns solely the credit score, without leaking any sensitive parts of the client's credit history. First, the function queries the client's bank(s) in order to retrieve their credit history. It then runs a credit score calculation on the accumulated credit history. Because the PFE has been vetted by the client a priori, they can ensure that no sensitive parts of their credit history are leaked.

Listing 4: The credit checker function. It first accumulates the history from multiple sources and then executes a check on the history.

```
import db

client_name   = argv[0]
client_secret = argv[1]
client_banks  = argv[2].split(',')

history = []

for bank in client_banks:
  res = db.call_on_peer(bank, 'credit', 'get_history',
                  [client_name, client_secret])
  history.append(res)

balance = 0
history = sorted(history)

for time, change in history:
  balance += change

  if balance < 0:
      return False

return True
```

Listing 4 shows a simplified version of the credit checker. After vetting, the client will call this function and pass along a list of their banks, as well as a secret that is only shared between the banks and the client. The secret serves as a certificate that the function has been vetted by the client. After executing PFEs on the client's banks, it will check the client's overall balance. In a real-world system, this check would be replaced by a more sophisticated mechanism.

### 4.4 Confidential Elections

Using computers for voting applications is non-trivial, because of the unique characteristics of elections. Elections must be both private, as in nobody's votes shall be revealed, and verifiable, as in one can check that each person has voted at most once. TeeVote is a voting service that stores its data in CreDB to achieve ensure confidentiality and integrity. Clients are able to securely cast ballots and verify that their vote has been counted. However, they are not able to see other people's vote, except for the final tally. The application uses two collections communities and votes.

Communities track membership of users and are used to determine the quorum for a vote. A vote is associated with a community. The community defines who is allowed to take part in the vote and what the quorum is for it to pass. The technical reasons for a requiring a quorum are two-fold: First, if the number of participants is too low, one might be able to trace back the voting behavior of a specific user. Second, there is no source of trusted time, so TEEvote will allow casting votes until a quorum has been reached.

Participants interact with TEEvote through a client-side application that directly connects to the CreDB instance. When starting the application for the first time, it will generate a public-private key pair associated with that user. It will then use the key pair to establish its secure connection and identity with the CreDB instance. Through this secure connection, clients can issue votes, query election results, and create new communities.

Policies enforce that only well-formed community and vote objects can be created. Well-formed in this context means that a person can only join the same community at most once. Further, votes must contain quorum information that matches the communities at the time of creation.

## 5 Experimental Evaluation

We evaluate the prototype on both macro and microbenchmarks. Note that evaluation is done under a single server setup and that the observed performance could be improved using horizontal or vertical [33] scaling. For all following experiments, the server process was hosted on Ubuntu 17.10 running Linux 4.13. The machine is equipped with 32GB of RAM and an Intel Core i7 6700K CPU offering 8 logical cores. The current prototype uses version 2.1 of the Intel SGX SDK and is compiled using GNU g++7. The client workload is distributed across multiple machines to make sure only the server's processing power can be a possible bottleneck. The main takeaway from the result in this section is that while the overheads associated with this kind of secure hardware are significant, they can be mitigated using efficient implementation and paging techniques.

## 5.1 Transactional Performance

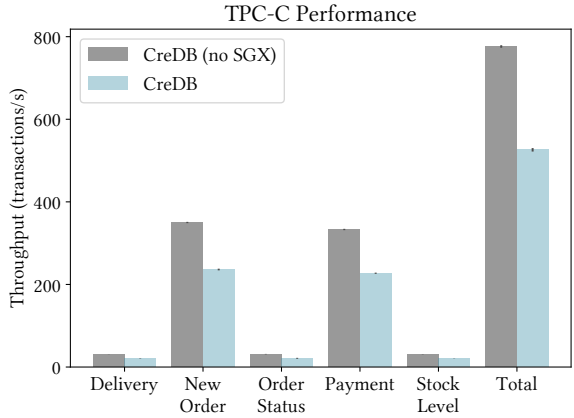

Figure 6: The performance of CreDB under a TPC-C workload compared to CreDB without SGX

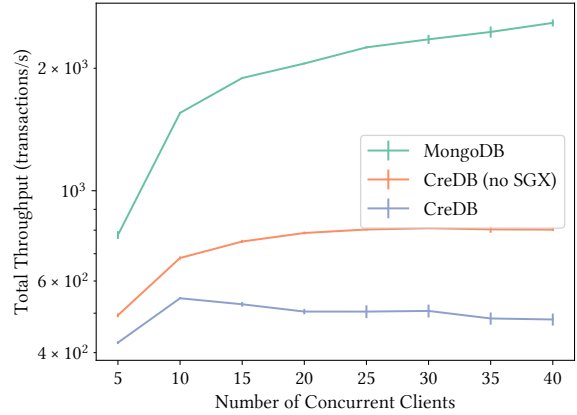

Figure 7: The transactional performance of CreDB under a changing number of clients compared to CreDB without SGX and MongoDB

We expose CreDB to a TPC-C workload and compare it to a version of CreDB that doesn't run in SGX. The experimental setup contains four warehouses and a dataset of about one gigabyte. Intel will most likely provide hardware with much larger EPC sizes in the future. We thus assume that the chosen dataset size is indicative of how future versions of CreDB will perform on larger datasets. We use py-tpcc[1], a Python implementation of TPC-C, for all measurements. For CreDB, data is stored normalized. In particular, each order is a distinct object and not part of the client's record.

Figure 6 visualizes the observed performance, broken down for each query type. Each setup was evaluated under the number of clients that yielded the highest overall throughput. As expected the version of CreDB that runs without a trusted environment performs significantly better.

We further plotted both systems over a changing number of clients in Figure 7 in order to visualize the impact of limited trusted memory. Additionally, we plot the performance of MongoDB under the same workload. We evaluated MongoDB on denormalized data, as we observed it to perform better than MongoDB on normalized data. In these experiments, MongoDB uses the WiredTiger storage engine writing to a memory-mapped location. Note that MongoDB has no support for ACID transactions and its performance is therefore not degraded by transaction aborts. Additionally, MongoDB does not provide the strong integrity guarantees and does not annotate the ledger with transactional information. The comparison to MongoDB solely serves as an ideal baseline.

We observe that each system scales up with an increasing number of clients. However, CreDB's throughput quickly reaches its peak of about 500tx/s. We pinpoint this limitation to the fact that once the EPC memory size is exhausted, threads will start competing for memory. The variant of CreDB without SGX yields in about twice the performance until concurrent transactions become the main bottleneck. MongoDB outperforms both implementations of CreDB due to the differences in application semantics described in the previous section.

---
[1] https://github.com/apavlo/py-tpcc

## 5.2 Microbenchmarks

**Trusted Execution** In order to understand the performance impact of the TEE, we compare CreDB against MongoDB as well as the version without SGX in microbenchmarks. As shown in Figure 7, even with the previously described optimizations, the current implementation suffers a significant loss in performance. Some of this performance loss stems from the fact that data has to be encrypted by the SGX environment. Figure 8a shows CreDB with and without SGX in two different situations: A read-only workload and workload with mixed reads and writes. Both workloads are small enough to fit into EPC memory. We observe that CreDB has about a 30% overhead compared to its counterpart that runs without SGX.

**Security Policies** An important question is how much the execution of security policies harm performance. To evaluate this, we store a 1kb object on the server. We then measure the latency of ten clients concurrently accessing this object either with a policy protecting the access or not. The policy we implemented is identical to that in Listing 1. Figure 8b shows a CDF of the latencies per I/O operation in each case. We observe that even with a security policy in place, operations are still executed in less than a millisecond. A similar performance impact was observed for other stored procedures.

**Witness Generation** Creating a signed certificate of a set of events incurs a non-trivial overhead. In Figure 8c we evaluate CreDB on a workload with read-only transactions, once while

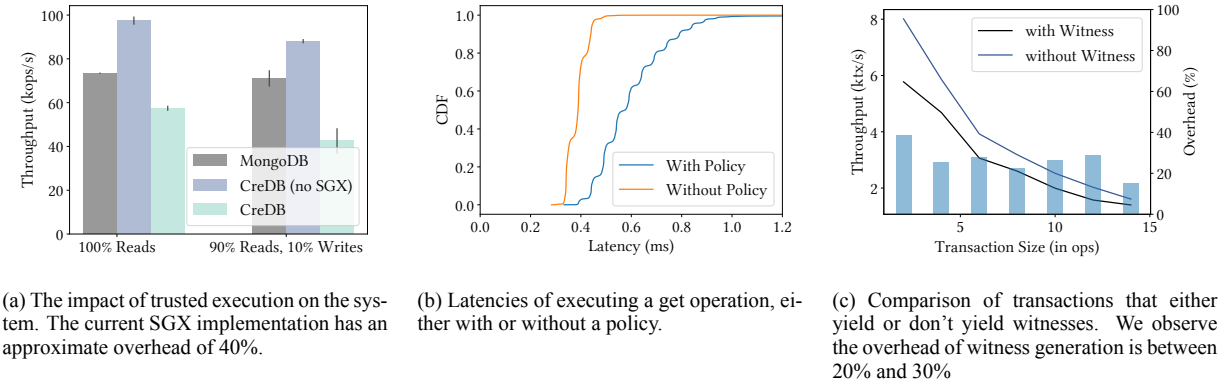

(a) The impact of trusted execution on the system. The current SGX implementation has an approximate overhead of 40%.

(b) Latencies of executing a get operation, either with or without a policy.

(c) Comparison of transactions that either yield or don't yield witnesses. We observe the overhead of witness generation is between 20% and 30% generating witnesses and once while not generating any witnesses. We observe that the witness generation produces roughly the same overhead while decreasing a little with larger transaction sizes. We attribute this overhead to the comparatively expensive asymmetric cryptographic operation that is needed to create the witness signature. While larger transactions have to sign a larger witness, they still only perform one cryptographic operation which explains why the overhead decreases with larger transactions.

Figure 8: Microbenchmark results

## 6 Discussion and Future Work

**More expressive object semantics** The evaluated prototype is built to scale with large-scale datasets but does not accommodate large individual objects. In particular, large objects are not able to fit in enclave memory and cannot be efficiently sharded into multiple blocks. CreDB currently mitigates this problem by supporting collection wide policies, which allow to break down application logic into multiple objects that are all part of the same collection. CreDB could further be extended by supporting large objects by only partially loading an object's content into memory. To make such an implementation efficient, the log would be required to only record the changes made to, and not the full values of, objects in order to keep log size small.

Another common problem in databases that is not addressed in this paper is schema consolidation, where different parties may not structure their stored objects in the same manner. We envision a simple type system where objects, both locally and remote, can be checked against a specification. This can then be used to not only check the object for a specific structure but also to verify its policy. Similarly, specifications can check other functions that are part of the object and speed up PFE, by defining a common protocol between multiple nodes.

**Handling Crash Failures** Crash failures can be caused by either broken hardware or malicious database operators that terminate the execution of the enclave. While the confidentiality of the hardware enclave cannot be broken this way, it may render the datastore inaccessible. In a real-world system, we assume that database operators are economically incentivized to replicate the encrypted data of the enclave sufficiently. Further, witnesses provide a way to retain facts about the datastore even after it went offline. SGX only provides limited support to protect stale data from being loaded back into the enclave after a crash, however, mechanisms such as ROTE [34] can ensure data freshness.

While out of the scope of this paper, a logical extension to the CreDB database model is to include replication requirements into the storage policies. For instance, nodes may advertise that they replicate all updates to a set of connected nodes. A write will only be considered successful when the write has been replicated to all members of the replica set. Such successful replication could then be certified by a witness similarly to other database operations.

**Side-channel Attacks** Side-channel attacks, that is attacks that observe the application's behavior through non-standard communication, such as looking at its CPU or cache usage, are of constant interest in the security community. Thus, several papers have addressed how the confidentiality of trusted hardware enclaves can be broken using such attacks [43]. Most of these attacks benefit from the fact that weak cryptographic code, e.g. where application secrets modify the control flow, is executed inside the enclave. While preventing CreDB nodes from side-channel attacks is beyond the scope of the paper, all cryptographic code in the enclave is implemented using constant-time libraries. We reserve addressing side-channel attacks against protected function evaluation for future work.

## 7 Related Work

**Enforcing Policies on Data** TEEs are one specific instance of secure hardware. The Nexus Operating System and its associated authorization logic [46] allow enforcing policies on applications. The Nexus uses a TPM to ensure only an authenticated and trusted kernel can be booted. The resulting operating system can then enforce complex constraints on applications running in its userspace. Traditional TPMs, such as the one used by Nexus, are usually harder to deploy as they require a fully trusted computation stack. TEEs, on the other hand, provide a "reverse sandbox" that shield enclave code from potentially malicious host operating systems. Thus, CreDB only requires trust in the trusted hardware and enclave code.

Information flow control is another common technique to enforce data policies on a programs execution. SIF [12] and Fable [47] use a combination of static and dynamic information flow tracking to enforce policies through compiler and runtime. Fabric [31] extends this paradigm to the distributed setting. While such techniques can protect data from malicious code, it cannot protect from other attackers, and is, thus, orthogonal to the mechanisms described in this paper.

In a distributed setting, certain Byzantine fault-tolerant consensus protocols can be used to shield a system from misbehaving principals [11, 9, 36]. In such an environment, the trust lies in the network itself and a large fraction of nodes behaving honestly. These protocols have been adopted in permissioned blockchains. Thus, they require careful selection of committee members and require a higher number of replicas than the approach described in this paper. Further, they do not shield from data leakage and cannot enforce access controls.

**Ensuring Data Integrity** Tamper-evident logs allow detecting Byzantine behaviors of storage servers [27, 53] and more complex applications [19]. While most of these mechanisms only provide fork consistency, A2M [13] uses trusted hardware to achieve strong consistency in such a setup. However, even if an audit mechanism provides strong consistency, to ensure detection of misbehavior, it requires that clients are honest and communicate with each other. Further, misbehavior can be detected only after the fact which is not a strong enough guarantee for many applications.

TrInc [25] provides a monotonic incrementer implemented in trusted hardware. Systems like CreDB can benefit from TrInc as it helps to protect from staleness attack, for example after an enclave is restarted. TrInc can also be used as a primitive to build Byzantine fault-tolerant systems with fewer replicas. However, it cannot be used to enforce access control to data.

Proof of Retrievability (PoR) [22, 45] allows verifying that a remote service indeed holds a dataset. PoR assumes a single client and thus is not suitable for some of CreDB's use cases. One possible application for PoR in CreDB would be to verify replication of encrypted data on a third party.

Concerto [5] is a datastore that achieves strong consistency using server-side integrity verification Due to batch verification, this approach achieves much higher performance than other mechanisms [26]. However, Concerto ensures only data integrity and does not guard the data from unwanted accesses.

Guardat [49] shields data from malicious applications by enforcing policies in the storage layer. The high-level motivation of Guardat is to reduce the attack surface of a complex system to a single policy-enforcing service. CreDB takes this concept a step further and enforces policies using trusted hardware.

**Encrypted Databases** If policy enforcement is not a requirement, i.e. users trust each other, operating on encrypted data might be sufficient to achieve confidentiality. Maheshwari et al. [32] presented one of the first encrypted databases. Their system stores hashes of the encrypted data in a small trusted hardware module to protect from tampering.

CryptDB [40] and Monomi [48] rely on homomorphic encryption of data. To make such a scheme efficient CryptDB does not encrypt all data and only supports a subset of the SQL language. TrustedDB [7] and Cipherbase [4] overcome this limitation by running queries on encrypted data using a trusted hardware module. All of these systems, to our knowledge, assume a single trusted client, potentially running multiple application. In contrast, the policy enforcement and accountability features in CreDB are designed with multiple distrusting clients in mind.

**Protecting Applications and Data using TEEs** Previous work demonstrated how to run mostly unmodified applications in trusted virtual machines [8] or containers [6] executing in a TEE. On a high level, the main difference between these systems and CreDB is the choice of abstraction. CreDB can provide applications with a trusted storage system, without having to execute the application itself fully in a trusted environment. Systems build on top of the CreDB API may thus yield in higher performance, with the tradeoff that the application has to be ported to this new API.

Ryoan [21] explores protecting data that is processed in the cloud using TEEs. This is achieved through a network of enclaves, similar to how CreDB nodes connect to each other. While Ryoan provides a sophisticated protected function evaluation mechanisms, it does not provide a mean for tamper-proof storage of data. Further, Ryoan only supports configurations that are static and does not provide mechanisms to scale or reconfigure the topology.

EnclaveDB [41] provides a mechanism similar to Cipherbase and TrustedDB but based implementation based on Intel SGX. Similar to our evaluation the system yields much better performance than conventional TPMs. EnclaveDB is currently limited to a set of clients and transactions specified at compile time of the transaction. Further, to our knowledge, it does not support federation of database nodes or timeline inspection.

**Cryptocurrencies and Blockchains** Digital payment systems have been researched for about two decades [42, 50]. However, Bitcoin [38] was the first system that found widespread use and interested outside out the academic community. Recently, the focus has shifted from using blockchains simply to enable digital payment towards more generalized high-integrity storage solutions. Permissioned [10, 3, 2], as well as permissionless [38, 52], blockchains require computationally expensive replication to ensure integrity. CreDB, on the other hand, presents itself as a standalone "blockchain of one" that does not require any form of replication for integrity.

Further TEEs can be used to build more energy-efficient consensus mechanisms. Proof of Elapsed Time (PoET) [1] uses a trusted hardware enclave to ensure nodes can only forge new blocks once they have waited a certain period of time. Milutinovic et al. [37] present a scheme that uses the random function in SGX instead of keeping track of elapsed time. Zhang et al. further improve this scheme by using the Proof of Work for useful computation [54]. These approaches are mostly suitable for the permissionless setting, where replicas can join and leave at any time. This is significantly different from the model CreDB provides.

## 8 Conclusion

We presented CreDB, a novel datastore that provides strong integrity guarantees through trusted execution environments. We classify CreDB as a "blockchain of one". This novel class of datastores combines the best of both world from conventional key-value stores and blockchains: they do not rely on massive replication or expensive proof of work, but still, enforce high integrity and assurance on the data.

This paper demonstrated that CreDB allows building high integrity distributed applications with relatively low effort. Our evaluation shows that this approach can handle hundreds of complex transactions a second on a single node. Unlike conventional BFT-systems, CreDB natively allows to scale up the system by creating a network of database nodes. We conclude that CreDB's design yields high performance compared to state-of-the-art permissioned and permissionless blockchains.

| Basic Operations | |
|---|---|
| get(C,k) | Get the most recent value of k |
| put(C,k,v) | Update the value of k |
| add(C,k,v) | Increment value of k by v |
| remove(C, k) | Delete object k |
| find(C, $\sigma$) | Find all object matching the predicate $\sigma$ |
| init_tx() | Yield a new transaction |
| commit(tx) | Commits a transaction |
| **Efficient Data Processing** | |
| create_index(C,n,f) | Create an index $n$ on the fields f |
| remove_index(C,n) | Remove the index $n$ |
| set_trigger(C,fx) | Invoke function $fx$ whenever the collection $C$ is modified |
| unset_trigger(C) | Unset the trigger on $C$ |
| **Reasoning about history** | |
| open(w1,w2,...) | Create a local timeline using a set of witnesses. |
| version_no(C,k) | Get the most recent version number of an object |
| get_history(C,k) | Get the history (previous versions and event ids) of an object |
| get_version(C,k,v) | Get the version v of an object |
| get_read_set(eid) | If $eid$ is part of a transaction, get the values read |
| get_write_set(eid) | If $eid$ part of a transaction, get other values written besides this one |
| get_op_source(eid) | Returns the party that caused the modification of this object |
| order(w1, w2) | Get the order of two witnesses |
| order(eid1, eid2) | Get the order of two events |
| **Inter-node connection** | |
| peer(addr) | Connect the CreDB instance to a remote instance |
| list_peers() | List all connected remote parties |
| **Function evaluation** | |
| call(C, k) | Call a stored procedure |
| execute(func) | Ship $func$ to the server and execute it |
| call_on_peer(p, k) | Call a function on remote party $p$ |

Table 2: List of all CreDB API calls. Operations are bound to a specific *collection* (or table) if C is their first argument.